\documentclass[onecolumn,prb,showpacs]{revtex4}
\usepackage{graphicx,psfrag}
\usepackage{slashed,amssymb,amsmath} 
\def\no{\noindent}
\newcommand{\nn}{\nonumber}

\begin{document}

\title{
Conductivity of disordered 2d binodal Dirac electron gas: \\Effect of internode scattering
}

\author{Andreas Sinner and Klaus Ziegler}
\affiliation{Institut f\"ur Physik, Universit\"at Augsburg\\D-86135 Augsburg, Germany}

\date{\today}

\begin{abstract}
We study the dc conductivity of a weakly disordered 2d Dirac electron gas with two bands and two spectral nodes,
employing a field theoretical version of the Kubo--Greenwood conductivity formula. In this paper we are concerned with the question how the internode scattering affects the conductivity. We use and compare two established techniques for treating the disorder scattering: The perturbation theory, there ladder and maximally crossed diagrams are summed up, and the functional integral approach. Both turn out to be entirely equivalent. For a large number of random potential configurations we have found only two different conductivity scenarios. Both scenarios appear independently of whether the disorder does or does not create the internode scattering. In particular we do not confirm the conjecture that the internode scattering tends to Anderson localization. 

\end{abstract}


\maketitle

\section{introduction}

Transport in two-dimensional (2d) electronic systems has been a subject of intense research for several decades.
One of the reasons for the attractiveness of this field is that quantum interference is strong in 2d and 
interesting phenomena emerge, such as the quantum Hall effect. Despite of its long history, some aspects of
electronic transport are still puzzling. 
For some time there was a consensus about the role disorder plays in transport processes, dominated
by Anderson localization of electronic wave functions for conventional 2d systems~[\onlinecite{Abrahams1979,Gorkov1979,Hikami1980,Vollhardt1980}].
A first hint, though, for an unconventional behavior was the transition between Hall plateaux in quantum Hall systems, which
indicated the existence of a metallic state in a 2d electronic system under special conditions~[\onlinecite{Hanein1998}].
Even more important was the discovery of metallic states in graphene [\onlinecite{Geim2005,kim07,elias09}] and in a number of 
chemical compounds, which are commonly referred to as topological insulators~[\onlinecite{Allen2010,Chen2012,Hasan2010,Qi2011}],
where the band structure has nodes and the electronic dispersion is linear in the vicinity of these nodes.
Although these compounds represent pristine 2d systems, they reveal a finite dc conductivity which is very robust against
thermal fluctuations and disorder. In the course of subsequent years these systems underwent careful studies from both 
the experimental and the theoretical point of view which clearly indicate that the finite dc value is a robust property. 

\vspace{1mm}
\no
The theoretical approach to the dc conductivity of disordered electron gases represents a notoriously difficult problem. The presence of disorder breaks explicitly the translational invariance, which obscures most of the tools of pristine microscopic analysis. An alternative way follows via the restoration of the translational invariance by averaging the observable quantity of interest over all disorder configuration. In practice though, the averaging of that sort can only be performed under assumption of a weak disorder, which guarantees for a well formed saddle-like shape of the free energy functional. Consequently, the calculations can be performed either in the Hamiltonian framework, which implies a partial summation of perturbative series in powers of random potential~[\onlinecite{Gorkov1979,Hikami1980,Vollhardt1980,Altshuler1981,Altshuler1994,Efetov1997,Levitov,Shon1998,
Ando2002,McCann2006,Khveshchenko2006,Tkachov2011,Schmeltzer2013}], or in Lagrangian framework, where the structure of elementary excitations around the saddle point in  a functional integral and the related field theoretical non--linear sigma models become the main object of studies~[\onlinecite{Lee1993,Wegner1979a,Wegner1979,Wegner1980,McKane1981,Hikami1981,Fradkin1986,Ziegler1997,Ziegler2009,Ziegler2012}]. 
In the course of several years both approaches developed themselves into independent branches of physics and are seldom compared directly with each other. Besides a detailed analysis of the transport properties of systems with a nodal spectrum, the comparison of different
approximative approaches is a central goal in this paper. 

\vspace{1mm}
\no
This paper aims at the understanding of the effects of different scattering types in systems with a nodal
electronic spectrum at chemical neutrality. Previous studies are inconclusive on the role intra- and internode scattering processes may play in electronic transport, ranging from essentially no difference~[\onlinecite{Shon1998}] to strong statements about a localization--antilocalization transition~[\onlinecite{McCann2006,Ando2002,Khveshchenko2006}], while general symmetry arguments indicate that massless modes should survive in the presence of random internode scattering~[\onlinecite{PhysicaE71,WSPaper2016}]. 

\vspace{1mm}
\no
We pursue a two--stage program, starting with the discussion of the averaging procedure based on the combination of functional integrals for quasiparticles with different statistics. This leads us to an effective field theory
in terms of graded matrix fields. The nontrivial vacuum of this theory turns out to be identical with the 
self--consistent Born approximation which is employed in the later parts of the paper. The inverse propagator 
of the effective action, obtained as Gaussian fluctuations around this vacuum,
turns out to be identical, up to a unitary transformation, with the solutions of the Bethe--Salpeter 
equation.
In the second part of the program we develop further the weak scattering formalism, formulated in our recent  papers~[\onlinecite{PhysicaE71,WSPaper2016,PhysRevB84,PhysRevB89,PhysRevB83,PhysRevB81,PhysRevB90}]. 
Besides the already mentioned comparison of two approaches, we calculate the conductivity for 16 different disorder
types. We detect two apparently different conductivity scenarios, which we call Type I and II scenario. 
The conductivity in first scenario arises solely from the channel of ladder diagrams (LC--channel), which includes all
diagrams without intersecting impurity lines, while the contribution from the channel of maximally crossed diagrams 
(MC--channel) is zero. In the second case, the MC-channel gives rise to the conductivity, while the contribution from the LC-channel is zero. Nonetheless, the evaluation of the  Bethe--Salpeter equation with the self--consistent evaluation of the scattering rate reveals the presence of massless modes in both channels for nearly all disorder types but with different degeneracy. This indicates that not all massless modes contribute to the Kubo--Greenwood formula.
Our analysis does not confirm the widespread perception that the discrete symmetries of the Hamiltonian determines
uniquely the transport properties.
In particular we do not observe any unique correlations between the Cartan class of a particular random Hamiltonian with its dc conductivity calculated from the Kubo--Greenwood formula. This is in line with previous
observations~[\onlinecite{Bernard2002,Bernard2012}].

\vspace{1mm}
\no
The scope of this paper is as follows: In Section~\ref{sec:model} we introduce the effective model which is studied
subsequently.
Apart from the  tight--binding approximation for non--interacting electrons on a 2d bipartite lattice, which yields a degenerated linear Dirac spectrum in the low--energy domain, we  discuss the phenomenology of different types of scattering in a random potential landscape. Furthermore we introduce and explain the conductivity Kubo--Greenwood 
formula based on the density--density correlation
function~[\onlinecite{Ludwig1994,PhysRevB83,Wegner1979,PhysRevB84,PhysRevB89,PhysRevB86,PhysRevB81,PhysRevB90}], 
which is evaluated in the course of this work. The link between this version of the Kubo--Greenwood formula and
diffusion is briefly discussed in Appendix~\ref{app:diff}. In Section~\ref{sec:path} we study the Gaussian 
fluctuations of the model within a functional integral approach, based on the notion of an intrinsic graded symmetry 
of the two--particle Green's function. Some technical details are moved into Appendices~\ref{app:det} and \ref{app:Sym}.  We continue in Section~\ref{sec:WSA} and in Appendix~\ref{app:PT}, where we approach the conductivity of the model by employing the
perturbative averaging technique.  We obtain generic expressions for the conductivity regardless of the disorder
type and evaluate it for 16 different types in Section~\ref{sec:Example}. In Section~\ref{sec:Discussion} we discuss the obtained results in comparison to some alternative approaches.

\section{Model}
\label{sec:model}

The dynamics of free electrons on bipartite lattices is governed by the family of tight--binding Hamiltonians 
\begin{equation}
\label{eq:TBH} 
H^{}_0 = -t\sum_{\langle rr^\prime\rangle}~(c^\dag_r d^{}_{r^\prime} + d^\dag_{r^\prime} c^{}_{r} )
\ ,
\end{equation}
where $c$ and $d$ ($c^\dagger$ and $d^\dagger$) are the fermionic annihilation (creation) operators with respect to each sublattice of a bipartite lattice, respectively. The neighboring lattice sites $r$ and $r^\prime$ are connected with the hopping amplitude $t$, and the summation is performed over nearest neighbor pairs only. Although we ultimately consider the case of the honeycomb lattice with the Fermi energy laying at the nodal points, the analysis is valid for other types of bipartite lattices as well. The Hamiltonian is readily diagonalized by Fourier transformation and yields on a 
honeycomb lattice a well--known spectrum with two nodes. As a striking feature, close to these nodes 
the fermion dispersion is linear and therefore describes massless Dirac particles~[\onlinecite{Wallace1947,Semenoff1984}]. In order for the Hamiltonian to remain invariant
under the time--reversal transformation both Dirac cones must be linked to each other by a parity 
transformation, i.e. they have different chiralities. Sometimes, the gas of such electrons at chemical neutrality is called the Weyl semimetal~[\onlinecite{Burkov2015}]. The effective low--energy Hamiltonian accounts for 
both Dirac species and reads~[\onlinecite{Wallace1947,Semenoff1984}] 
\begin{equation}
\label{eq:Hamiltonian}
H^{}_0 \simeq 
v_F\begin{pmatrix}
p^{}_1\sigma^{}_1+p^{}_2\sigma^{}_2 & & 0 \\
\\
0 & &  p^{}_1\sigma^{}_1-p^{}_2\sigma^{}_2
\end{pmatrix}
\equiv v_F\begin{pmatrix}
h & 0 \\
0 & \sigma_3 h^T\sigma_3 \\
\end{pmatrix}
= v_F(p^{}_1\Sigma^{}_{01} + p^{}_2\Sigma^{}_{32})
\ ,
\end{equation}
where matrices $\Sigma^{}_{ij} = \sigma^{}_i\otimes\sigma^{}_j$, $\sigma^{}_{i,j}$ are the unity ($i,j=0$) and 3 Pauli ($i,j=1,2,3$) matrices in the usual representation, the momentum operators are $p^{}_i = -i\hbar \nabla^{}_i$, 
$v_F$ denoting the Fermi velocity. 
Below we use dimensionless energy units, i.e. $\hbar v^{}_F=1$ in units of inverse length. 

\vspace{1mm}
\no
Disorder appears in the model in form of one--particle random potentials $V$ which are matrices in the Dirac space. 
Due to the mentioned nodal degeneracy in the electronic spectrum, 
all possible disorder scattering processes can be roughly subdivided into those with and without mixing
of electrons with different chiralities. 
The former processes are referred to as the intranode scattering. Corresponding random low--energy  
Hamiltonians have the diagonal block structure 
\begin{equation}
H = \left( 
\begin{array}{cc}
 h + V^{}_+ & 0 \\
 0 & \sigma_3 h^{T}\sigma_3 + V^{}_-
\end{array}
\right)
\ ,
\label{ham222}
\end{equation}
where 
$V^{}_\pm$ denotes the random part, i.e. each block can be diagonalized independently.
On the other hand, electrons whose scattering involves a change of the chirality are referred to as 
internode scattering processes. Corresponding Hamiltonians have the following generic form
\begin{equation}
H = \left( 
\begin{array}{cc}
 h^{}_+ & V \\
 V^\ast & \sigma_3h^{T}_+\sigma_3  
\end{array}
\right)
\ .
\end{equation}
The Hamiltonian can also be written as $H=H_0+V$ with the $4\times4$ random matrix 
\begin{equation}
V = v^{}_{ij}\Sigma^{}_{ij},
\end{equation}
where $v^{}_{ij}$ is the random coordinate-dependent part and the summation convention is used.  
Then, the decomposition coefficients with $i=0,3$ are those responsible for the intra-node scattering, 
while $i=1,2$ is for inter-node scattering. In this paper we concentrate on a somewhat simpler model, 
where each disorder type has a decomposition with one term only; i.e., 
\begin{equation}
V =  v\Sigma^{}_{ij}, \hspace{2mm} i,j\;\; {\rm fixed}.
\end{equation}
In a usual fashion we assume for the random part of the disorder potential a Gaussian distribution
independently for each site with
\begin{equation}
\label{eq:Correlator}
\langle v(r)\rangle^{}_g = 0,\hspace{5mm}  
\langle v(r) v(r^\prime) \rangle^{}_g = g\delta(r-r^\prime),
\end{equation}
where $g$ is referred to as the disorder strength and $\langle\cdots\rangle^{}_g$ denotes the ensemble averaging.
Then the disorder strength is measured in units of $(\hbar v^{}_F)^2$ and the quantity $s=(\hbar v^{}_F)^2$ 
represents an appropriate reference scale. 

\vspace{1mm}
\no
{\it Special cases:}
The case $V_+=-V_-=\mu\sigma_0$ (i.e., for $\Sigma_{30}$ disorder) is equivalent (up to an orthogonal transformation) 
to a single Dirac with random scalar potential~[\onlinecite{PhysRevB86}]. Moreover,
the case of $\Sigma_{03}$ is related to the random gap model~[\onlinecite{Ziegler2009}]. 
The latter has only one massless mode, corresponding to a massless
fermion mode in the functional integral~[\onlinecite{Ziegler1997}] or to a maximally crossed diagram in the weak perturbation series~[\onlinecite{PhysicaE71}].

\subsection{The Kubo--Greenwood conductivity formula}
\label{sec:Kubo}

Motivated by the phenomenon of diffusion at large times ~[\onlinecite{McKane1981,Ziegler2012,Huang1963}] (cf. Appendix \ref{app:diff}), we consider the field theoretical version of the Kubo--Greenwood conductivity formula
\begin{equation}
\label{eq:KuboGen1} 
\bar\sigma^{}_{\mu\mu} = \frac{e^2}{h}\lim_{\epsilon\to0}\int_{}^{}dE~\Gamma^{}_\mu(E,\epsilon)~\frac{f(E)-f(E+i\epsilon)}{i\epsilon},
\end{equation}
where $f(E)$ is the Fermi function, $\epsilon$ is the zero--temperature Matsubara frequency, and the disorder averaged core function reads
\begin{equation}
\label{eq:KuboGen2}
\Gamma^{}_\mu(E,\epsilon) = \epsilon^2{\rm Tr}^{}_d\sum_r~r^2_\mu~\langle G^+(r,0)G^-(0,r)\rangle^{}_g,
\end{equation}
where $G^+(r,r^\prime)$ ($G^-(r,r^\prime)$) denotes the microscopic advanced (retarded) single--particle Green's function at energy $E$ analytically continued into the upper (lower) halfplane
\begin{equation}
\label{eq:green_func}
G^{\pm}(r,r^\prime) = \langle r| [E \pm i\epsilon + H^{}_0 + V ]^{-1}| r^\prime\rangle.  
\end{equation}
The operator ${\rm Tr}^{}_d$ denotes the trace operator on the space of $d$--dimensional Green's functions. At frequencies small as compared to the typical band width of the clean system and well below room temperature we can employ the usual approximation $f(E+i\epsilon) \sim f(E) +i\epsilon\delta(E)$, which trivializes the energy integral in Eq.~(\ref{eq:KuboGen1}). Then the conductivity formula becomes
\begin{eqnarray}
\label{eq:Kubo1}
\bar\sigma^{}_{\mu\mu} &=& \lim_{\epsilon\to0}\epsilon^2\left.\left(-\frac{\partial^2}{\partial q^2_\mu}\right)\right|_{q=0} \sum_r e^{iq\cdot r} 
\langle G^+_{ij}(r,0) G^-_{ji}(0,r)\rangle^{}_g,
\end{eqnarray}
where we have used the Fourier representation of the position operator and the summation convention for matrix elements with respect to the spinor index. Due to the randomness each realization of the system lacks translational invariance and the Green's function depends on both sites $r$ and $r'$. The establishing of the translational invariance is achieved by the averaging procedure. In Sections \ref{sec:path} and \ref{sec:WSA} we demonstrate how the averaging is performed within a functional--integral formalism and a diagrammatic weak--scattering formalism, respectively.

\section{Functional integral approach}
\label{sec:path}

The functional integral approach to disorder averaged quantities and eventually to the conductivity can be based
on a functional integral with a fermionic and a bosonic field. Usually the propagators for both fields
are identical, and the functional integral is called supersymmetric~[\onlinecite{Efetov1997}]. An alternative
to this supersymmetric functional integral is based on the identity between the 
determinants of two different Green's functions~[\onlinecite{Ziegler1997,Ziegler2009,Ziegler2012}]
\begin{equation}
\det G=\det G'
\ .
\end{equation}
Then we can use the Green's function $G$ for the fermionic field and $G'$ for the bosonic field (or vice versa)
and calculate the average product $\langle GG'\rangle$ within this functional-integral representation. 
In our specific case we have the determinant identity (cf. Appendix~\ref{app:det})
\begin{equation}
\label{eq:Det}
\det[i\epsilon+H^{}_0+V] = \det[i\epsilon+H^T_0-{\cal O}V{\cal O}]
\ ,
\end{equation}
where $H^T_0$ describes the Hamiltonian transposed on all spaces and the operator $\cal O$ satisfies the
condition 
\begin{equation}
\label{eq:ChirTr}
H^{}_0=-{\cal O}H^T_0{\cal O}.
\end{equation}
For the Dirac Hamiltonian it is important to notice the transposition rule of the momentum operator: 
$p^T_i=-p^{}_i$, due to the property of the differential operator
$\partial^T_{x_i}=\partial^{}_{-x^{}_i}=-\partial^{}_{x_i}$. 
Therefore with $H^{}_0=p^{}_1\Sigma^{}_{01} + p^{}_2\Sigma^{}_{32}$ follows
$H^T_0 = -p^{}_1\Sigma^{}_{01} + p^{}_2\Sigma^{}_{32}$ and either ${\cal O}=\Sigma^{}_{01}$ or 
${\cal O}=\Sigma^{}_{31}$. 
The identity~(\ref{eq:Det}) is valid for the Dirac Hamiltonian in Eq.~(\ref{eq:Hamiltonian}) and for any random 
potential $V=v\Sigma^{}_{ij}$ obeying ${\rm Tr}^{}_dV=0$. The latter condition rules out 
the totally degenerated random potential $V=v\Sigma^{}_{00}$, which is discussed separately in Appendix~\ref{app:Sym}.

\vspace{1mm}
\no
The identity~(\ref{eq:Det}) allows us to write the disorder averaged of two--particle 
Green's function in Eq.~(\ref{eq:Av2pGr})
\begin{eqnarray}
\nn
K^{}_{rr^\prime} &=& {\rm Tr}^{}_d \langle [i\epsilon + H^{}_0 + V]^{-1}_{rr^\prime} [-i\epsilon + H^{}_0 +V]^{-1}_{r^\prime r} \rangle^{}_g\\
\nn
&=& - {\rm Tr}^{}_d \langle [i\epsilon+H^{}_0+V]^{-1}_{rr^\prime} {\cal O}[i\epsilon + H^T_0 - {\cal O}V{\cal O}]^{-1}_{r^\prime r}{\cal O}\rangle^{}_g \\
\label{eq:IntCore}
&=&
{\cal O}^{}_{jk}{\cal O}^{}_{li}\langle\chi_{r,kj}{\bar\chi}_{r',il}\rangle^{}_{\hat Q}
\ .
\end{eqnarray}
For the last equation we have used a functional integral with respect to the matrix field
\begin{equation}
{\hat Q}=\begin{pmatrix}
Q & \chi \\
{\bar\chi} & iP
\end{pmatrix}
\ .
\end{equation}
The $4\times 4$ matrices $Q,P$ with commuting elements, and $\bar\chi,\chi$ with anticommuting elements are 
hermitian, and thus can be expanded in a basis which includes a four--dimensional unity matrix and 15 
traceless matrices, which are fifteen $\Gamma$--matrices. The functional integral is defined as
\begin{equation}
\langle\cdots\rangle^{}_{\hat Q} = \int\cdots e^{-{\cal S}^{}_G}\prod_r d\bar\chi_rd\chi_r dQ_r dP_r
\label{funct_int}
\end{equation}
with the action  ${\cal S}^{}_{G}$
\begin{eqnarray}
\label{eq:Gauss1}
{\cal S}^{}_{G}[\hat Q] &=& \frac{1}{2}{\rm Trg}\int d^2rd^2r^\prime\left\{
\frac{1}{g}\delta^{}_{rr^\prime}\hat Q^{}_r\hat Q^{}_{r^\prime}-
\hat G(r,r^\prime)\hat Q^{}_{r^\prime}\hat \Sigma\hat G(r^\prime,r)\hat Q^{}_r\hat\Sigma
\right\}
\end{eqnarray}
and with frequency-- and position--dependent Green's functions 
\begin{equation}
\hat G(r,r^\prime) = \langle r|[iz+\hat H^{}_0]^{-1}|r^\prime\rangle,
\end{equation}
where $\hat H^{}_0 = {\rm diag}[H^{}_0,H^T_0]$, $H^{}_0 = p^{}_1\Sigma^{}_{01} + p^{}_2\Sigma^{}_{32}$,  and $z=\epsilon+\eta$. The operator $\rm Trg$ denotes the graded trace with
${\rm Trg} {\hat Q}={\rm Tr}Q-i{\rm Tr}P$. Eq. (\ref{eq:Gauss1}) is the result of the saddle-point approximation of 
a more complex functional integral, taking only fluctuations up to quadratic (Gaussian) order into account. 
The saddle point equation reads
\begin{equation}
\label{eq:Saddle1}
\frac{i\eta}{g}  = - \int\frac{d^2q}{(2\pi)^2} ~ \left[ iz + \hat H_0 \right]^{-1} 
= \int\frac{d^2q}{(2\pi)^2}~\frac{iz}{z^2+q^2}.
\end{equation}
The action in Eq. (\ref{eq:Gauss1}) can be decomposed into bosonic and fermionic sectors:
${\cal S}_G={\cal S}[Q]+{\cal S}[P]+{\cal S}[\bar\chi,\chi]$. Expanding the kernel of the bilinear expression
in powers of the momentum of the field (equivalent to a gradient expansion in real space), we obtain 
at zeroth order
\begin{eqnarray}
\label{eq:Fluc1}
{\cal S}[Q] &=& \frac{1}{2g} \int\frac{d^2q}{(2\pi)^2}~Q_{ab}(q) M^Q_{a\alpha,b\beta}Q_{\alpha\beta}(-q), \\
\label{eq:Fluc2}
{\cal S}[P] &=& \frac{1}{2g} \int\frac{d^2q}{(2\pi)^2}~P_{ab}(q) M^P_{a\alpha,b\beta}P_{\alpha\beta}(-q),\\
\label{eq:Fluc3}
{\cal S}[\bar\chi,\chi] &=& \frac{1}{g}\int\frac{d^2q}{(2\pi)^2}~\bar\chi_{ab}(q)M^\chi_{a\alpha,b\beta}\chi_{\alpha\beta}(-q)
\end{eqnarray}
with
\begin{eqnarray}
\label{eq:Gauss2Q} 
M^Q_{a\alpha,b\beta} &=& \delta^{}_{\alpha b}\delta^{}_{a\beta} - g\int\frac{d^2p}{(2\pi)^2}~
\displaystyle[\Sigma^{}_BG^-_{}(p)]^{}_{\beta a}[\Sigma^{}_BG^-_{}(p)]^{}_{b\alpha},
\\
\label{eq:Gauss2P} 
M^P_{a\alpha,b\beta} &=& \delta^{}_{\alpha b}\delta^{}_{a\beta} - g\int\frac{d^2p}{(2\pi)^2}~
\displaystyle[{\cal O}\Sigma^{}_BG^+_{}(p){\cal O}]^{}_{\beta a}[{\cal O}\Sigma^{}_BG^+_{}(p){\cal O}]^{}_{b\alpha},
\\
\label{eq:Gauss2F} 
M^\chi_{a\alpha,b\beta} &=& \delta^{}_{\alpha b}\delta^{}_{a\beta}  - g\int\frac{d^2p}{(2\pi)^2}~
\displaystyle[\Sigma^{}_BG^-(p)]^{}_{\beta a}[{\cal O}\Sigma^{}_BG^+(p){\cal O}]^{}_{b\alpha}
\end{eqnarray}
with $G^\pm(p)=[\pm i\eta + H^{}_0]^{-1}$. Note the missing combinatorial factor 1/2 in ${\cal S}[\bar\chi,\chi]$,
since it counts both permutations $\bar\chi\chi$ and $\chi\bar\chi$.  Once the fluctuations in
Eq.~(\ref{eq:Fluc1}) -- (\ref{eq:Fluc3}) are expressed as quadratic forms in terms of real valued 
expansion coefficients of
fields $Q,P,\bar\chi$,and $\chi$,the eigenvalues of the corresponding inverse propagator matrices become nonnegative. 

\vspace{1mm}
\no
The expansion of Eq. (\ref{eq:Gauss1}) to second order in the momenta yields the inverse (diffusion) propagators.
This enables us to compare the functional-integral approach with a weak-scattering expansion of 
Section~\ref{sec:WSA}.
In particular, the expression for zero momentum, which represents the mass matrices, has non-zero eigenvalues.
Zero eigenvalues are most important, since they correspond to diffusion modes.

\section{Perturbative averaging approach to the conductivity}
\label{sec:WSA}

In the second part of our program we approach the dc conductivity within a weak scattering approach, in which the disorder average is performed perturbatively. This is the standard method in the weak--localization approach,
in which the Green's functions are expanded in terms of the random scattering elements and averaged 
afterwards~[\onlinecite{Gorkov1979,Hikami1980,Vollhardt1980,Altshuler1981,Altshuler1994,Efetov1997,Levitov,Shon1998,Ando2002,McCann2006,Khveshchenko2006,Tkachov2011,Schmeltzer2013}].
As an extension of the analysis in our recent papers~[\onlinecite{PhysicaE71,WSPaper2016}],
where we were concerned with the conductivity behavior in the single--cone approximation, 
here we take both cones into account. The averaged two-particle Green's function 
\begin{equation}
\label{eq:Av2pGr}
K^{}_{rr^\prime} = \langle G^+_{nj}(r,0) G^-_{jn}(0,r)\rangle^{}_g
\end{equation}
is treated within a perturbative expansion in powers of weak scattering rate $\eta$, which plays the role of the order parameter in  diffusive regime. It is determined self--consistently from 
\begin{equation}
\label{eq:SCBA}
\pm i\eta = -\frac{1}{d}\sum_{r^\prime}{\rm Tr}^{}_d~\langle V(r) G^\pm(r,r^\prime)V(r^\prime)\rangle^{}_g = 
-\frac{g}{d}{\rm Tr}^{}_d\left[\Sigma\bar G^\pm(r,r)\Sigma\right],
\end{equation}
where $\Sigma$ are disorder matrices in Dirac space. The averaged one--particle Green's functions become 
translationally invariant and read
\begin{equation}
\bar G^{\pm}(r,r^\prime) = \langle r| [\pm iz + H^{}_0 ]^{-1}| r^\prime\rangle = \int\frac{d^2p}{(2\pi)^2}~e^{ip(r-r^\prime)} 
\frac{\tilde H^{}_0\mp iz}{p^2+z^2},
\end{equation}
with $z=\epsilon+\eta$ and $\tilde H^{}_0$ denoting the Fourier--transformed Hamiltonian. $\tilde H^{}_0$ is 
not rotational invariant and does not contribute to the integral. We recognize then in Eq.~(\ref{eq:SCBA}) the 
saddle point of the functional integral defined in Eq.~(\ref{eq:Saddle1}).  For $\epsilon\sim0$ we get
\begin{equation}
\label{eq:SCBA2}
1 - \int\frac{d^2p}{(2\pi)^2}~\frac{g}{p^2+z^2} \sim \frac{\epsilon}{\eta} .
\end{equation}
Partial summations of all ladder (LC) and maximally crossed (MC) diagrams (see Appendix~\ref{app:PT}) lead to 
\begin{eqnarray}
\nn
\bar\sigma^{}_{\mu\mu} &=& \lim_{\epsilon\to0}~{\epsilon^2}\left.\left(-\frac{\partial^2}{\partial 
q^2_\mu}\right)\right|_{q=0}\sum_{rr^\prime}e^{iq\cdot r}~\bar G^+_{ij}(r^\prime,0)\bar G^-_{kn}(0,r^\prime)\\
\label{eq:ATPGF2}
&\times&\left[\left(1 -g[\bar G^+\Sigma][\bar G^-\Sigma]
\right)^{-1}_{rr^\prime|nj;ik} + \left(1-g[\bar G^+\Sigma][\bar G^{-}\Sigma]^{T}\right)^{-1}_{rr^\prime|nk;ij}\right],
\end{eqnarray}
where the full transposition operator $T$ applies to all degrees of freedom.
We shall call the terms in the second line of Eq.~(\ref{eq:ATPGF2})
\begin{eqnarray}
\label{eq:LC}
a)\hspace{1mm}
t= g[\bar G^+\Sigma][\bar G^-\Sigma] \ , \ \ 
b)\hspace{1mm}
\tau= g[\bar G^+\Sigma][\bar G^{-}\Sigma]^T \ \
\end{eqnarray}
the LC (Eq.~(\ref{eq:LC}a)) and the MC (Eq.~(\ref{eq:LC}b)) channel matrices, 
respectively~[\onlinecite{PhysicaE71}]. These matrices read in Fourier representation
\begin{eqnarray}
\label{weak_scatt1}
t^{}_{r^\prime r|ab;cd} &=& g\int\frac{d^2q}{(2\pi)^2}~e^{-iq\cdot(r-r^\prime)}
\int\frac{d^2p}{(2\pi)^2}~[\bar G^+(p)\Sigma]^{}_{ac}[\bar G^{-}(q+p)\Sigma]^{}_{bd} ,\\\
\label{weak_scatt2}
\tau^{}_{r^\prime r|ab;cd} &=& g\int\frac{d^2q}{(2\pi)^2}~e^{-iq\cdot(r^\prime-r)}
 \int\frac{d^2p}{(2\pi)^2}~[\bar G^+(p)\Sigma]^{}_{ac}[\bar G^{-}(q-p)\Sigma]_{db}.
\end{eqnarray}
The different signs in the argument of $t$ and $\tau$ are a consequence of the transposition on the position 
space in the MC--channel. 
Fourier transformed matrices $1-{\tilde t}_{q}$ and $1-{\tilde \tau}_{q}$ for
$\epsilon=0$ and $q=0$ read
\begin{eqnarray}
\label{eq:MLC}
M^{LC}_{ab;cd} &=& \left. {\delta^{}_{ac}\delta^{}_{bd}} - g\int\frac{d^2p}{(2\pi)^2}~
[\bar G^+(p)\Sigma]^{}_{ac}[\bar G^{-}(p)\Sigma]^{}_{bd} \right|_{\epsilon=0},\\
\label{eq:MMC}
M^{MC}_{ab;cd} &=& \left.{\delta^{}_{ac}\delta^{}_{bd}} - g\int\frac{d^2p}{(2\pi)^2}~[\bar G^+(p)\Sigma]^{}_{ac}
[\bar G^{-}(-p)\Sigma]_{db} \right|_{\epsilon=0}\ .
\end{eqnarray}
Matrices $M$ in Eqs.~(\ref{eq:MLC}) and (\ref{eq:MMC}) are diagonalized by the momentum independent orthogonal
transformations $O^{}_{}$. The eigenvalues of the matrices $M$ provide a decay length for the matrices in
Eq.~(\ref{eq:LC}). In particular, a vanishing (e.g. gapless) eigenvalue gives a  long-range diffusion-like 
behavior. Depending on the type of disorder there may or may not be gapless modes. If gapless modes exist, 
then for small momenta and frequencies we get in the LC channel (in the channel of MC diagrams analogously) 
\begin{eqnarray}
\label{eq:InvProp}
t^{}_{r^\prime r} \sim O^{}_{LC}~{\rm diag}[t]^{}_{r^\prime r}~O^T_{LC} &\sim&
\int\frac{d^2q}{(2\pi)^2}~e^{iq\cdot(r-r^\prime)} ~
O^{}_{LC}
\left[
\begin{array}{ccc}
\label{eq:InvPropA}
\displaystyle \left(\frac{\epsilon}{\eta} +gD^{}_0q^2\right)1^{}_N & & {\bf 0}^T \\
\\
{\bf 0}  & &  \hat\Lambda^{}_{D-N}
\end{array}
\right]
O^T_{LC},
\end{eqnarray}
where $N$ is the dimension of the subspace populated by gapless modes. $\hat\Lambda^{}_{D-N}$ represents 
the diagonal matrix with non--zero eigenvalues of $M^{LC}$ as matrix elements.
$D$ is the dimension of the full space on which the product $G^+\otimes G^-$ dwells, and $D^{}_0$ is the expansion 
coefficient, which fortunately turns out to be the same for all studied cases:
\begin{equation}
\label{eq:DiffCoef}
D^{}_0 = \frac{1}{2} \int\frac{d^2p}{(2\pi)^2}~\frac{1}{[p^2+\eta^2]^2}
\sim \frac{1}{8\pi \eta^2}
\ .
\end{equation}
Finally, $\bf 0$ is an $(D-N)\times N$ matrix whose elements are zero. Using the convolution formula for 
the tensor product of two Green's functions
\begin{equation}
\bar G^+_{ij}(r,0)\bar G^-_{kn}(0,r)
 = \int\frac{d^2p}{(2\pi)^2}~e^{-ip\cdot r}~\int\frac{d^2k}{(2\pi)^2}~\bar G^+_{ij}(k)\bar G^-_{kn}(k+p),
\end{equation}
we can bring Eq.~(\ref{eq:ATPGF2}) to the form where the derivatives with respect to the external momentum $q$ can be easily evaluated. 
After performing the dc limit we arrive at
\begin{eqnarray}
\nn
&\displaystyle 
\bar\sigma^{}_{\mu\mu} = 2g\eta^2 D^{}_0 \int\frac{d^2p}{(2\pi)^2}~\bar G^+_{ij}(p)\bar G^-_{kn}(p)\times&\\
\label{eq:Kubo3}
&\displaystyle
\left\{\left[
O^{}_{LC}\left(
\begin{array}{ccc}
\displaystyle  1^{}_{N^{}_{LC}} & & {\bf 0}^T \\
{\bf 0} & & \hat{\bf 0}_{D-N^{}_{LC}}
\end{array}
\right)
O^T_{LC}\right]^{}_{nj;ik} +
\left[
O^{}_{MC}\left(
\begin{array}{ccc}
\displaystyle  ~1^{}_{N^{}_{MC}} & & {\bf 0}^T \\
{\bf 0} & & \hat{\bf 0}^{}_{D-N^{}_{MC}}
\end{array}
\right)
O^T_{MC}\right]^{}_{nk;ij}
\right\},
&
\end{eqnarray}
where $\hat{\bf 0}^{}_{D-N^{}_{}}$ denotes a $(D-N)\times(D-N)$ zero matrix. This result indicates that the conductivity
$\bar\sigma^{}_{\mu\mu}$ is finite for a positive scattering rate $\eta$. In particular, it does not have a 
logarithmic divergence in the limit of zero temperature, as predicted in some weak localization calculations. 
For correct evaluation of the conductivity, the disorder strength parameter $g$ in front of the integral must be compensated with the help of the saddle point condition~(\ref{eq:SCBA2}):
\begin{equation}
\label{eq:rand}
 \int\frac{d^2p}{(2\pi)^2}~\bar G^+_{ij}(p)\bar G^-_{kn}(p)\Gamma^{}_{ikjn} = \frac{1}{g} + {\rm remaining\; terms},
\end{equation}
where the quantity $\Gamma$ denotes the full tensor in the curly brakets.
For ultimate values of the conductivity we must determine the orthogonal matrices $O^{}_{LC}$ and $O^{}_{MC}$. 
This requires the explicit knowledge of the type of disorder, i.e. matrices $\Sigma$. 
In Section~\ref{sec:Example} we discuss solutions of Eq.~(\ref{eq:Kubo3}) for a number of disorder types.

\begin{table}
\begin{center}
\begin{tabular}{||c||c|c|c|c|c|c||}
\hline\hline
\hspace{1mm} Matrix~$\Sigma^{}_{ab}$ \hspace{1mm}  & \hspace{4mm}{ 0}\hspace{4mm} & \hspace{4mm} a \hspace{4mm} & \hspace{4mm} 2a \hspace{4mm} & \hspace{4mm} 2b \hspace{4mm} &  \hspace{3mm} 2(a+b) \hspace{3mm} & \hspace{3mm} a+2b  \hspace{3mm}  \\
\hline\hline 
$\Sigma^{}_{00}$ &    4    &  8  &  4   &  0  &    0    &   0    \\
\hline 
$\Sigma^{}_{01}$ &    4    &  4  &  0   &  4  &    0    &   4    \\
\hline  
$\Sigma^{}_{02}$ &    2    &  4  &  2   &  2  &    2    &   4    \\
\hline  
$\Sigma^{}_{30}$ &    2    &  4  &  2   &  2  &    2    &   4    \\ 
\hline  
$\Sigma^{}_{31}$ &    2    &  4  &  2   &  2  &    2    &   4    \\
\hline   
$\Sigma^{}_{32}$ &    4    &  4  &  0   &  4  &    0    &   4    \\
\hline  
$\Sigma^{}_{10}$ &    2    &  4  &  2   &  2  &    2    &   4    \\
\hline  
$\Sigma^{}_{11}$ &    2    &  4  &  2   &  2  &    2    &   4    \\ 
\hline  
$\Sigma^{}_{13}$ &    2    &  4  &  2   &  2  &    2    &   4    \\
\hline 
$\Sigma^{}_{20}$ &    2    &  4  &  2   &  2  &    2    &   4    \\
\hline  
$\Sigma^{}_{21}$ &    2    &  4  &  2   &  2  &    2    &   4    \\
\hline  
$\Sigma^{}_{23}$ &    2    &  4  &  2   &  2  &    2    &   4    \\ 
\hline
\end{tabular}
\caption{Eigenvalues and their number of the mass matrices in LC channel for the random potentials $v\Sigma^{}_{ab}$ giving Type I conductivity.}
\end{center}
\begin{center}
\begin{tabular}{||c||c|c|c|c|c|c||}
\hline
\hspace{1mm} Matrix~$\Sigma^{}_{ab}$ \hspace{1mm} & \hspace{4mm}{ 0}\hspace{4mm} & \hspace{4mm} a \hspace{4mm} & \hspace{4mm} 2a \hspace{4mm} & \hspace{4mm} 2b \hspace{4mm} &  \hspace{3mm} 2(a+b) \hspace{3mm} & \hspace{3mm} a+2b  \hspace{3mm}  \\
\hline\hline 
$\Sigma^{}_{03}$ &    2    &  4  &   2  &  2  &    2     &    4    \\
\hline 
$\Sigma^{}_{33}$ &    0    &  8  &   0  &  4  &    4     &    0    \\
\hline  
$\Sigma^{}_{12}$ &    2    &  4  &   2  &  2  &    2     &    4    \\
\hline  
$\Sigma^{}_{22}$ &    2    &  4  &   2  &  2  &    2     &    4    \\ 
\hline\hline
\end{tabular}
\caption{Eigenvalues and their number of the mass matrices in LC channel for the random potentials $v\Sigma^{}_{ab}$ giving Type II conductivity.}
\end{center}
\end{table}

\begin{table}
\begin{center}
\begin{tabular}{||c||c|c|c|c|c|c||}
\hline\hline
\hspace{1mm} Matrix~$\Sigma^{}_{ab}$ \hspace{1mm} & \hspace{4mm}{ 0}\hspace{4mm} & \hspace{4mm} a \hspace{4mm} & \hspace{4mm} 2a \hspace{4mm} & \hspace{4mm} 2b \hspace{4mm} &  \hspace{3mm} 2(a+b) \hspace{3mm} & \hspace{3mm} a+2b  \hspace{3mm}  \\
\hline\hline 
$\Sigma^{}_{00}$ &    4    &  8  &   4  &  0  &  0    &   0   \\
\hline 
$\Sigma^{}_{01}$ &    0    &  4  &   4  &  0  &  4    &   4   \\
\hline  
$\Sigma^{}_{02}$ &    2    &  4  &   2  &  2  &  2    &   4   \\
\hline  
$\Sigma^{}_{30}$ &    2    &  4  &   2  &  2  &  2    &   4   \\ 
\hline  
$\Sigma^{}_{31}$ &    2    &  4  &   2  &  2  &  2    &   4   \\
\hline   
$\Sigma^{}_{32}$ &    0    &  4  &   4  &  0  &  4    &   4   \\
\hline  
$\Sigma^{}_{10}$ &    2    &  4  &   2  &  2  &  2    &   4   \\
\hline  
$\Sigma^{}_{11}$ &    2    &  4  &   2  &  2  &  2    &   4   \\ 
\hline  
$\Sigma^{}_{13}$ &    2    &  4  &   2  &  2  &  2    &   4   \\
\hline 
$\Sigma^{}_{20}$ &    2    &  4  &   2  &  2  &  2    &   4   \\
\hline  
$\Sigma^{}_{21}$ &    2    &  4  &   2  &  2  &  2    &   4   \\
\hline  
$\Sigma^{}_{23}$ &    2    &  4  &   2  &  2  &  2    &   4   \\ 
\hline
\end{tabular}
\caption{Eigenvalues and their number of the mass matrices in MC channel for the random potentials $v\Sigma^{}_{ab}$ giving Type I conductivity.}
\end{center}
\begin{center}
\begin{tabular}{||c||c|c|c|c|c|c||}
\hline\hline
\hspace{1mm} Matrix~$\Sigma^{}_{ab}$ \hspace{1mm} & \hspace{4mm}{ 0}\hspace{4mm} & \hspace{4mm} a \hspace{4mm} & \hspace{4mm} 2a \hspace{4mm} & \hspace{4mm} 2b \hspace{4mm} &  \hspace{3mm} 2(a+b) \hspace{3mm} & \hspace{3mm} a+2b  \hspace{3mm}  \\
\hline\hline 
 $\Sigma^{}_{03}$ &   2    &  4  &   2  &  2  &    2     &   4   \\
\hline 
 $\Sigma^{}_{33}$ &   4    &  0  &   4  &  0  &    0     &   8   \\
\hline  
 $\Sigma^{}_{12}$ &   2    &  4  &   2  &  2  &    2     &   4   \\
\hline  
 $\Sigma^{}_{22}$ &   2    &  4  &   2  &  2  &    2     &   4   \\ 
\hline
\end{tabular}
\caption{Eigenvalues and their number of the mass matrices in MC channel for the random potentials $v\Sigma^{}_{ab}$ giving Type II conductivity.}
\end{center}
\end{table}

\section{Conductivity for different random potentials} 
\label{sec:Example}

Now we can evaluate the conductivity of a binodal Dirac electron gas in a random potential from 
Eq.~(\ref{eq:ATPGF2}). In order to investigate the role of the disorder mediated intermixing 
of Dirac electrons with different chiralities we use the full renormalized  Dirac propagators 
$\bar G^\pm(p)$ which read in Fourier representation
\begin{equation}
\bar G^\pm(p) = \frac{1}{p^2+\eta^2} \left(
\begin{array}{cccc}
 \mp i\eta  & p^{}_1 - i p^{}_2 & 0 & 0 \\
 \\
 p^{}_1 + i p^{}_2  & \mp i\eta & 0 & 0 \\
 \\
 0 & 0 & \mp i\eta & p^{}_1+i p^{}_2 \\
 \\
 0 & 0 & p^{}_1 -ip^{}_2 & \mp i\eta 
\end{array}
\right).
\end{equation}
Different disorder types were studied by inserting respective disorder matrices $\Sigma^{}_{ij}$ directly into Eqs.~(\ref{eq:ATPGF2}) and then unfolding them up to the form of Eq.~(\ref{eq:Kubo3}). In all we have evaluated
the conductivity for 16 different disorder configurations. 
Technically, the evaluation of the  conductivity contributions for each disorder type does not differ much from the
single--cone model evaluation presented in our recent papers Ref.~[\onlinecite{PhysicaE71,WSPaper2016}]. 
The number of massless modes is obtained by counting  zero eigenvalues of the corresponding mass matrices given
in Eqs.~(\ref{eq:MLC}) and (\ref{eq:MMC}). Despite high dimensionality of the matrices (16$\times$16 or even
64$\times$64), the order of the characteristic polynomial is small, since the matrices are sparsely occupied. 
In the case of 16$\times$16--matrices, of 256 matrix elements only 24 to 32, depending on the disorder type, 
are non--zero. Moreover the mass matrices are symmetric. Consequently, they can be diagonalized analytically
using a computer algebra package. Then, the calculation of the conductivity from Eq.~(\ref{eq:Kubo3}) can be
carried out exactly. Surprisingly, for all 16 types of random matrices we have discovered only two distinct 
conductivity types. Both types occur for disorder potentials with and without chirality mixing and have
their own features:

\vspace{2mm}
\no
{\bf Type 1:} Reveals a total suppression of the contribution from the MC--channel: 
\begin{eqnarray}
\sigma^{MC}_{\mu\mu} \sim 0.
\end{eqnarray}
The conductivity is entirely controlled by the LC--channel, which yields
\begin{eqnarray}
\sigma^{LC}_{\mu\mu} \sim 4\eta^2D^{}_0\sim \frac{1}{2\pi}
\ .
\end{eqnarray}
This conductivity type was found for 
$
\Sigma^{}_{00}, \Sigma^{}_{01}, \Sigma^{}_{02}, \Sigma^{}_{30}, \Sigma^{}_{31}, \Sigma^{}_{32}, 
\Sigma^{}_{10}, \Sigma^{}_{11}, \Sigma^{}_{13}, \Sigma^{}_{20}, \Sigma^{}_{21}, \Sigma^{}_{23}. 
$
\vskip0.2cm

\no
{\bf Type 2:} Demonstrates a picture somewhat reciprocal to the Type 1: The conductivity contribution from the LC--channel is fully suppressed
\begin{eqnarray}
\sigma^{LC}_{\mu\mu} \sim 0,
\end{eqnarray}
while the MC--channel gives 
\begin{eqnarray}
\sigma^{MC}_{\mu\mu} \sim 8\eta^2D^{}_0(1 - a),
\end{eqnarray}
where the parameter $a$ is defined below in Eq.~(\ref{eq:sorthand}). The Type 2 conductivity exhibits a logarithmically suppressed conductivity and was found for $\Sigma^{}_{03}, \Sigma^{}_{33}, \Sigma^{}_{12}, \Sigma^{}_{22}$. 

\vspace{1mm}
\no
Having found only two conductivity types, it is natural to expect that the corresponding disorder matrices share
some common features. We would expect that the microscopic random Hamiltonians, which lead to the same conductivity
type, belong to the same Cartan classes. This property of the random Hamiltonians is usually brought into connection
with the expected Anderson localization transition in the 
system~[\onlinecite{Zirnbauer1996,Altland1997,Bernard2002,Bernard2012}]. According to the Altland--Zirnbauer classification of $N\times N$ random matrices, the behavior of a random Hamiltonian under the time--reversal, 
particle--hole, and chiral symmetry transformations determines to which Cartan class this Hamiltonian belongs:
\begin{eqnarray}
\label{eq:CC1}
T H^\ast T^\dag = H,  & & {\rm Time-reversal\;\; symmetry}\\
\label{eq:CC2}
CH^TC^\dag = - H,  & & {\rm Particle-hole\;\; symmetry} \\ 
\label{eq:CC3}
PHP^\dag = - H, & & {\rm Chiral\;\; symmetry}
\ .
\end{eqnarray}
Here $H^\ast$ denotes the complex conjugate of the Hamiltonian $H$ and $H^T$ its transposition on all spaces. 
If we extend the $N\times N$ random matrix ensembles by including a spatial Dirac operator, the number of classes
increases at least to 17 for the single--node Dirac--like Hamiltonians~[\onlinecite{Bernard2002,Bernard2012}]
and may increase even further when two nodes and inter--node scattering is included. Therefore, the
classification of random Hamiltonians given by Eq.~(\ref{ham222}) remains a great challenge.
In the present work we focus only on the characterization of the transport properties.
We notice that the mass matrices in both LC and MC channels are unique to each disorder matrix $\Sigma^{}_{ij}$.
The composition of the sets of eigenvalues for each disorder matrix $\Sigma^{}_{ij}$ (in particular the number
of zero eigenvalues) may differ from each other, even if the microscopic Hamiltonians formally belong to the
same Cartan class. The sets of eigenvalues of the mass matrices are shown in Tables~I-IV for both channels
with the notation  
\begin{equation}
a=\int\frac{d^2p}{(2\pi)^2}~\frac{gp^2}{[p^2+\eta^2]^2}, \,\,\, b=\int\frac{d^2p}{(2\pi)^2}~\frac{g\eta^2}{[p^2+\eta^2]^2}.
\label{eq:sorthand}
\end{equation}
A systematic property of the eigenvalue sets consists in the total number of zero eigenvalues per LC and MC channels
together. There are always 4 with the only important exclusion of the totally degenerated disorder which 
couples to the unity matrix on the extended space  $\Sigma^{}_{00}$. In the latter case the number of zero 
eigenvalues is doubled; i.e., 8 instead of 4. This observation is linked back to the continuous symmetry on the configuration space which spontaneously breaks down in order to give rise to the gapless excitations is the same for all disorder types~[\onlinecite{Ziegler2012,PhysRevB86,PhysicaE71}]. 
In the case of $\Sigma^{}_{00}$--case the symmetry lives in a space which is twice as large, cf. Appendix~\ref{app:Sym}. This explains
the higher number of massless modes. The order parameter in all cases is the scattering rate $\eta$. 

\vspace{1mm}
\no
Within the functional--integral approach we obtain the same sets of eigenvalues for the various disorder types, as  
summarized in Tables I-IV. A direct correspondence between the modes of LC-- and MC--channel on the one hand,
and the fermionic and the bosonic field on the other can be obtained for the random matrices with $\Sigma^{}_{01}$ 
and $\Sigma^{}_{32}$. These cases are related to the Type 1 conductivity (cf. Tables I and II). 
For $\Sigma^{}_{33}$, which is Type 2 conductivity (Tables III and IV), the full set of eigenvalues in
LC--channel (with no zero eigenvalues) is reproduced in both bosonic ($Q$ and $P$) channels, while the full
set of eigenvalues of MC--channel arises from the fermionic channel. Thus, the full equivalency of both averaging procedures is established. 

\vspace{1mm}
\no
Since both conductivity types occur due to the spontaneous breaking of the same continuous symmetry, the
origin of the different transport properties must be in the discrete transformations of the effective Hamiltonian
on the configuration space. These symmetries are difficult to realize within the perturbative approach of Section~\ref{sec:WSA} but can be recognized using functional integral representation which we discussed in Section~\ref{sec:path}. The generator $\cal T$ with ${\cal T}{\cal T}=1$ can be considered as an effective time--reversal transformation, when it distinguishes Hamiltonian $H_0$ and the random part $V$ through the following properties: 
\begin{equation}
\label{eq:DisSym}
a)\; H^{}_0={\cal T}H^T_0{\cal T},\;\; 
b)\; V=-{\cal T}{\cal O}V{\cal O}{\cal T}
\ .
\end{equation}
Eq.~(\ref{eq:DisSym}a) is solved either for ${\cal T}=\Sigma^{}_{02}$ or ${\cal T}=\Sigma^{}_{32}$. 
Combining this with the chiral symmetry condition of Eq.~(\ref{eq:ChirTr}) for the full Hamiltonian we 
make the following general observation: If a matrix $\Sigma^{}_{ij}$ obeys both equalities
\begin{equation}
\label{eq:DisSym2}
a)\;\Sigma^{}_{ij} = {\cal T}{\cal O}\Sigma^{}_{ij}{\cal O}{\cal T},\,\,\, {\rm and}\,\,\,b)\;\Sigma^{}_{ij} = - {\cal O}\Sigma^{}_{ij}{\cal O},
\end{equation}
for ${\cal O}=\Sigma^{}_{01}$, ${\cal T}=\Sigma^{}_{32}$, we have
\begin{equation}
{\cal T}(H_0^T+{\cal O}V{\cal O}){\cal T}=(H_0+V)
\ .
\label{transformation3}
\end{equation}
This is valid for all matrices in $V$ which lead to the Type~2 conductivity, (i.e., for $\Sigma^{}_{03}, \Sigma^{}_{33}, \Sigma^{}_{12}, \Sigma^{}_{22}$), which therefore all disobey Eq.~(\ref{eq:DisSym}b) while satisfying Eq.~(\ref{eq:DisSym2}b). 
On the other hand, none of the corresponding matrices in $V$ for the Type~1 scenario (i.e., for $\Sigma^{}_{00}, \Sigma^{}_{01}, \Sigma^{}_{02}, \Sigma^{}_{30}, \Sigma^{}_{31}, \Sigma^{}_{32}, \Sigma^{}_{10}, \Sigma^{}_{11}, \Sigma^{}_{13}, \Sigma^{}_{20}, \Sigma^{}_{21}, \Sigma^{}_{23} $)
fulfill Eqs.~(\ref{eq:DisSym2}) simultaneously. Thus, the relations (\ref{eq:DisSym2})
characterize Type 2 conductivity, the absence of them characterizes Type 1 conductivity.

\section{Discussion}
\label{sec:Discussion}

The vast body of literature which exists on the subject of the disordered Dirac electron gas raises
a natural question of how results presented in this paper are related to previous theoretical 
and experimental findings.  Particularly the weak-localization 
approach~[\onlinecite{Gorkov1979,Hikami1980,Shon1998,McCann2006,Tkachov2011}] 
deserves special attention. It is commonly believed to represent the fastest and most natural way
to describe the transport in conventional disordered metals. In its essence it allows to compute so-called 
perturbative quantum corrections to the classical Drude-Boltzmann conductivity and appears in the form of 
infinite series of Feynman diagrams. Clearly, such an expansion is possible only in the parametric 
region close to the classical regime; i.e., for weak quantum fluctuations. 
A contribution to the Kubo formula comes from the velocity--velocity correlation function
\begin{equation}
\label{eq:WLKubo}
\sigma^{}_{\mu\mu} \sim {\rm tr}~ v^{}_\mu G^{}_{+}v^{}_\mu G^{}_{-},
\end{equation}
where $G^{}_+(G^{}_-)$ represents the advanced (retarded) Green's function. It is dominant
in Kubo formula of a one-band system~[\onlinecite{Levitov}] and accounts for the most important 
corrections to the conductivity.  In particular, the corrections arising by summing up the so--called
maximally crossed diagrams, shown schematically in appendix in Fig.~\ref{fig:Diagrams}.
The absence of an extended Fermi surface in the half-filled nodal systems has led to the 
incorrect perception that the 'ladder'-approximation would break down for the nodal systems. 
Consequently, in the overwhelming majority of investigations which employ the weak-localization 
approach, the Dirac electron gas is considered to be doped; i.e., 
a chemical potential is introduced to create a Fermi circle instead of the Fermi node, 
which makes it similar to conventional metals. Precisely at the node the reduced Kubo formula in Eq.~(\ref{eq:WLKubo}) has its limitations, cf. Appendix~\ref{app:WLKubo}.

However, a closer inspection of the Dirac node reveals that in the Dyson-Schwinger (or self-consistent Born)
approximation disorder generates an exponentially small momentum scale
$\sim a^{-1}\exp(-2\pi/g)$ for the leading term of the perturbation series
~[\onlinecite{Lee1993,Wegner1979a,McKane1981,Fradkin1986}]. This solves the problem of the spurious
and unphysical logarithmic singularities found in the expansion around the trivial vacuum with
a vanishing density of states of Ref.~[\onlinecite{Aleiner2006}], which was also discussed
previously for the band edges in a one-band model some time ago [\onlinecite{oppermann79,ziegler83}].
In other words, the self-consistent approximation provides a non-trivial vacuum state with non-zero density of states
around which a perturbative expansions appears with finite non-singular terms. 
The onset of that scale in the density of states for weak disorder is clearly seen 
numerically ~[\onlinecite{Huckestein2003,Pereira2006,Wu2008}] but also, which is even more important, 
experimentally~[\onlinecite{kim07}]. 

In summary, our approach, based on the expansion around a non-trivial vacuum,
reproduces correctly the experimental transport results, whereas the expansion around the vacuum with a vanishing density of states
[\onlinecite{Aleiner2006}] clearly contradicts the latter with a divergent conductivity.
This divergency needs to be cut off by an additional artificially introduced scattering
time. This step is rather confusing, since even in the classical Boltzmann approach the elastic
random scattering creates a finite conductivity. 


Fortunately, through the discovery of graphene it became possible to test the various theoretical predictions with experimental measurements.  
In the past our approach~[\onlinecite{Ziegler1997}] has been compared with experimental work, including the seminal article on graphene 
by Novoselov et al. on graphene~[\onlinecite{Geim2005}] and showed good agreement. More recently we also compared our  
theoretical results for finite-size scaling with experimental results~[\onlinecite{WSPaper2016}], and found excellent agreement. 
Moreover, we have used two different approaches in this article, namely a functional integral approach based on symmetry considerations (Sect. \ref{sec:path})
and a perturbative approach based on the Bethe-Salpeter equation (Sect. \ref{sec:WSA}), and the results of both approaches agree.

\section{Conclusions}

Despite decades of intense research several aspects of the disorder physics are still challenging. Novel low--dimensional materials defy traditional views and reveal a new and hitherto unexpected physics. The dc conductivity of 2d disordered nodal electron gases represents a striking example of that kind. Following the traditional theories it must necessarily disappear in  infinite systems no matter how weak the disorder is. Yet the experimental evidence clearly speaks against this perception. Numerous attempts to reconcile those facts ultimately made the internode scattering processes responsible for localization. According to this hypothesis, once such processes are realized in the physical systems, it must inevitably become an insulator. 

\vspace{2mm}
\no
In this work we performed an extensive studies of the dc conductivity of chemically neutral disordered 2d binodal Dirac 
or Weyl electron gas aiming at the question, how the internodal scattering affects the transport in such systems. 
Using the version of the Kubo--Greenwood formula based on the density--density correlation function, we compared 
two technically very different approaches to the disorder averaging: the perturbative weak scattering technique and the 
non--perturbative functional integral approaches. For a large number of disorder potentials we obtained from both 
approaches the same diffusion propagators, which eventually give rise to the conductivity. Our findings reveal a very 
simple picture of what happens in those systems. There are only two distinct conductivity types, the one which yields a 
universal conductivity value in infinite systems and the second which reveals a logarithmically suppressed conductivity. 
However, in contrast for instance to the weak--localization theory, even in the latter case Anderson localization never 
occurs and the system remains a conductor. The internode scattering can be definitely ruled out as the primary cause of 
the localization in nodal systems, since both conductivity types are observed for both inter- and intranode scattering 
potentials. A similar statement can also be made for the wide spread perception on the role of discrete symmetries of 
the microscopic Hamiltonians. While they definitely are important as we demonstrate in the present work, there is no 
direct connection between the Cartan classification of the microscopic random Hamiltonian and the localization
properties. At this stage, the true reason why the transport follows different types for different 
random potentials remains inconclusive.

\section*{ACKNOWLEDGMENTS}

This work was supported by a grant of the Julian Schwinger Foundation for Physical Research.

\appendix

\section{Diffusion and conductivity}
\label{app:diff}

Starting from the transition probability
\begin{equation} 
P_{rr'}(t)=\sum_{j,j'}\langle|\langle r,j|e^{-iHt}|r',j'\rangle|^2\rangle^{}_g,
\end{equation}
with $\langle\cdots\rangle^{}_g$ denoting the disorder averaging we can calculate the diffusion coefficient as
\begin{equation} 
D=\lim_{\epsilon\to 0}\epsilon^2\sum_r (r_k-r_k')^2 \int_0^\infty dt~P_{rr'}(t) e^{-\epsilon t}
\ .
\end{equation}
With the Green's function $G(z)=(H-z)^{-1}$, we can write for the integral
\begin{equation} 
\int_0^\infty dt~P_{rr'}(t) e^{-\epsilon t}
=\int dE~{\rm Tr}^{}_d\left\{G_{rr'}(E+i\epsilon)\left[
G_{r'r}(E-i\epsilon) -G_{r'r}(E+i\epsilon)\right]\right\},
\end{equation}
where ${\rm Tr}^{}_d$ is the trace with respect to the spinor index. Only the contribution with poles on both sides
of the real axis is relevant. Then we can write
\begin{equation} 
D=\lim_{\epsilon\to 0}\epsilon^2\sum_r (r_k-r_k')^2 
\int dE~ {\rm Tr}^{}_d\left[G_{rr'}(E+i\epsilon)G_{r'r}(E-i\epsilon)\right].
\end{equation}

\section{Proof of Eq.~(\ref{eq:Det})}
\label{app:det}

Our aim is to give a proof for Eq.~(\ref{eq:Det})
\begin{equation}
\label{eq:app:det1}
\det[i\epsilon 1 + H^{}_0 + V] = \det[i\epsilon 1 + H^T_0 -{\cal O}V{\cal O}],
\end{equation}
for $H^{T}_0 = -{\cal O}H^{}_0{\cal O}$, ${\cal O}{\cal O}=1$ and ${\rm Tr}^{}_d[H^{}_0+V]=0$, that is 
\begin{equation}
\label{eq:app:det2} 
H^T_0-{\cal O}V{\cal O} = -{\cal O}[H^{}_0+V]{\cal O}
\end{equation}
Be $\lambda^{}_i$ and $\mu^{}_i$ eigenvalues of matrices $i\epsilon 1 + H^{}_0 + V$ and $i\epsilon 1 + H^T_0 -{\cal O}V{\cal O}$, respectively, then Eq.~(\ref{eq:app:det1}) is equivalent to 
\begin{equation}
\label{eq:app:det3}
\prod^{2N}_{i=1}\lambda^{}_{i} = \prod^{2N}_{i=1}\mu^{}_{i}.
\end{equation}
The eigenvalues in turn follow from the solutions of the corresponding secular equations
\begin{equation}
\det[(\lambda-i\epsilon) 1 - H^{}_0-V] = 0 = \det[(\mu-i\epsilon) 1 - H^{T}_0+{\cal O}V{\cal O}],
\end{equation}
such that $\lambda^\prime = \lambda - i\epsilon$ and $\mu^\prime = \mu - i\epsilon$ represent the eigenvalues of random Hamiltonians $H^{}_0 + V$ and $H^T_0 -{\cal O}V{\cal O}$, respectively, and therefore real numbers. Then with Eq.~(\ref{eq:app:det2}) follows
\begin{equation}
\det[\lambda^\prime 1-H^{}_0-V] = 0 =  \det[\mu^\prime 1 -H^T_0 + {\cal O}V{\cal O}] = \det[{\cal O}(\mu^\prime 1+H^{}_0+V){\cal O}],
\end{equation}
i.e. 
\begin{equation}
\det[\lambda^\prime 1-H^{}_0-V] = 0 = \det[\mu^\prime 1+H^{}_0+V],
\end{equation}
which obviosly implyes $\lambda^\prime_i = -\mu^\prime_i$. Because random Hamiltonians are traceless, the eigenvalues of d dimensional matrices  always appear in bundles, each of which sums up to zero separately, e.g. in d=2 there are only two with opposite sing; in our case d=4 there are two pair with opposite sing $\lambda^{\prime}_i \to \pm|\lambda^\prime_i|$, i.e. $\mu^\prime_i\to\mp|\lambda^\prime_i|$. 
Then Eq.~(\ref{eq:app:det3}) becomes
$$
\prod^{N}_{i=1}(i\epsilon + |\lambda^{}_{i}|)(i\epsilon - |\lambda^{}_{i}|)= \prod^{N}_{i=1}(i\epsilon - |\lambda^{}_{i}|)(i\epsilon + |\lambda^{}_{i}|),
$$
which obviously proofs Eq.~(\ref{eq:Det}).

\section{Embedding of the totally degenerated random potential}
\label{app:Sym}

In order to imbed the totally degenerated disorder which couples to the matrix $\Sigma^{}_{00}$ into the functional integral formalism of Section~\ref{sec:path} we use the fact that $\langle V^{2n+1}\rangle^{}_g=0$ for any integer $n$, i.e. the sign of the random potential in the Hamiltonian does not affect the ultimate results. Then we can write the two--particles Green's function as
\begin{eqnarray}
\nn
K^{}_{rr^\prime} &=&  \frac{1}{2}{\rm Tr}^{}_4 \langle [i\epsilon + H^{}_0 + V]^{-1}_{rr^\prime}[-i\epsilon + H^{}_0 + V]^{-1}_{r^\prime r} \rangle^{}_g +
\frac{1}{2}{\rm Tr}^{}_4\langle [i\epsilon + H^{}_0 - V]^{-1}_{rr^\prime}[-i\epsilon + H^{}_0 - V]^{-1}_{r^\prime r} \rangle^{}_g \\
&=& \frac{1}{2}{\rm Tr}^{}_8 \langle
\left[ 
\begin{array}{ccc}
i\epsilon + H^{}_0 + V & & 0 \\
\\
0 & & i\epsilon + H^{}_0 - V 
\end{array}
\right]^{-1}_{rr^\prime}
\left[ 
\begin{array}{ccc}
-i\epsilon + H^{}_0 + V & & 0 \\
\\
0 & & -i\epsilon + H^{}_0 - V 
\end{array}
\right]^{-1}_{r^\prime r}
\rangle^{}_g, 
\end{eqnarray}
and carry out calculations in this extended representation.

\section{Dyson and Bethe--Salpeter equations in weak scattering regime} 
\label{app:PT}

We briefly recapitulate the derivation of the disorder averaged Green's function for the case of weak scattering. To keep the notation simpler, here we use the symbol $\langle\cdots\rangle$ to denote the ensemble average. We start with the expansion of the microscopic Green's function into the geometric series:
\begin{eqnarray}
\label{eq:Dyson1}
G = [G^{-1}_0+V]^{-1} \sim G^{ }_0 - G^{}_0VG^{}_0 + G^{}_0VG^{}_0VG^{}_0 - G^{}_0VG^{}_0VG^{}_0VG^{}_0  \cdots
\end{eqnarray}
Applying the averaging procedure in accord with Eq.~(\ref{eq:Correlator}) we may drop all terms containing odd powers of $V$. Then rearranging we get:
\begin{eqnarray}
\nn
\langle G\rangle &=& \langle G^{}_0 + G^{}_0VG^{}_0VG^{}_0 + G^{}_0VG^{}_0VG^{}_0VG^{}_0VG^{}_0  \cdots \rangle \\
\nn
&=& \langle G^{}_0 + G^{}_0VG^{}_0V (G^{}_0 + G^{}_0VG^{}_0VG^{}_0 + G^{}_0VG^{}_0VG^{}_0VG^{}_0VG^{}_0  \cdots) \rangle \\
\label{eq:Dison2}
&\sim&  G^{}_0 + G^{}_0\langle VG^{}_0V\rangle\langle G\rangle = G^{}_0(1 + \langle VG^{}_0V\rangle\langle G \rangle),
\end{eqnarray}
where we made use of the weak scattering conjecture. This yields the renormalization of $G$:
\begin{equation}
\label{eq:Dison3}
\langle G \rangle \sim \left(G^{-1}_0 - \langle VG^{}_0V\rangle\right)^{-1} \sim \bar G.
\end{equation}
In an analogous fashion we consider  the two--particle Green's function:
\begin{eqnarray}
\langle G^+ G^- \rangle = \langle \left([G^+_0]^{-1}+V\right)^{-1} \left([G^-_0]^{-1}+V\right)^{-1}\rangle.
\end{eqnarray}
We expand the Green's function into the geometric series:
\begin{eqnarray}
\nn
\langle G^+ G^- \rangle &=& \langle\left(G^+_0 - G^+_0VG^+_0 + G^+_0VG^+_0VG^+_0 \cdots\right)
\left(G^-_0 - G^-_0VG^-_0 + G^-_0VG^-_0VG^-_0 \cdots\right)\rangle\\
\label{eq:Bethe1}
&=& G^+_0 G^-_0 +\langle (G^+_0VG^+)(G^-_0VG^-)\rangle -G^+_0G^-_0\langle VG^-\rangle - \langle G^+V \rangle G^+_0G^-_0.
\end{eqnarray}
For assumed weak scattering the last two terms are of the order zero and we may retain only first two terms which allow for self--consistency
\begin{equation}
\label{eq:Bethe2}
\langle G^+G^-\rangle \sim G^+_0G^-_0 + \langle (G^+_0VG^+)(G^-_0VG^-)\rangle .
\end{equation}
Expanding it in a matrix basis and exploiting the weak scattering conjecture we furthermore have
\begin{eqnarray}
\nn
\langle G^+_{ij} G^-_{kl}\rangle &\sim& G^+_{0,ij}G^-_{0,kl} + \langle G^+_{0,ia}V^{}_{ab}G^+_{bj}G^-_{0,k\alpha}V^{}_{\alpha\beta}G^-_{\beta l}\rangle \\
&\sim& G^+_{0,ij}G^-_{0,kl} +\langle G^+_{0,ia}V^{}_{ab}G^-_{0,k\alpha}V^{}_{\alpha\beta}\rangle \langle G^+_{bj}G^-_{\beta l}\rangle,
\end{eqnarray}
which then reversibly rewritten reads
\begin{equation}
G^+_{0,ij}G^-_{0,kl} \sim \left[\delta^{}_{ib}\delta^{}_{k\beta} -\langle G^+_{0,ia}V^{}_{ab}G^-_{0,k\alpha}V^{}_{\alpha\beta} \rangle\right] 
\langle G^+_{bj}G^-_{\beta l} \rangle = \langle[1 -  (G^+_0V)(G^-_0V)]\rangle^{}_{ik;b\beta}\langle G^+_{bj}G^-_{\beta l}\rangle.
\end{equation}
The expression on the left hand side is invariant under the averaging procedure, i.e. we can rewrite the last equation as follows:
\begin{equation}
\langle\left[1-(G^+_0V)(G^-_0V)\right]^{}_{ik;b\beta} \left([1-(G^+_0V)(G^-_0V)]^{-1}_{b\beta; nm} G^+_{0,nj}G^-_{0,ml}
- \langle G^+_{bj}G^-_{\beta l}\rangle \right) \rangle=0.
\end{equation}
Making again use of the weak scattering conjecture we obtain the Bethe--Salpeter equation
\begin{equation}
\label{eq:BetheSalpeter1}
\langle G^+_{nj} G^-_{ml} \rangle \sim\langle[1-(G^+_0V)(G^-_0V)]^{-1}\rangle^{}_{nm;ik}G^+_{0,ij}G^-_{0,kl} 
\end{equation}
which is the equation we have to evaluate. Expanding the disorder matrix in (\ref{eq:BetheSalpeter1}) into the geometric series we get
\begin{eqnarray}
\nn
\langle[1-(G^+_0V)(G^-_0V)]^{-1}\rangle^{}_{nm;ik} &=& \delta^{}_{ni}\delta^{}_{mk} +\langle (G^+_{0}V)^{}_{ni}(G^-_{0}V)^{}_{mk} \rangle \\
&+& \langle  (G^+_{0}V)^{}_{n\alpha} (G^-_{0}V)^{}_{m\beta} (G^+_{0}V)^{}_{\alpha i}(G^-_{0}V)^{}_{\beta k}
\rangle + \cdots.
\end{eqnarray}
Then performing the average in accord with Eq.~(\ref{eq:Correlator}) by means of the Wick theorem we get 
\begin{eqnarray}
\nn
\langle[1-(G^+_0V)(G^-_0V)]^{-1}\rangle^{}_{nm;ik} &\sim&  \delta^{}_{ni}\delta^{}_{mk}  + 
g (G^+_{0}\Sigma)^{}_{ni}(G^-_{0}\Sigma)^{}_{mk} + \\
\label{eq:Wick2}
&+&g^2[(G^+_{0}\Sigma)^{}_{n\alpha}(G^+_{0}\Sigma)^{}_{\alpha i}][(G^-_{0}\Sigma)^{}_{m\beta}(G^-_{0}\Sigma)^{}_{\beta k}]\\
\label{eq:Wick1}
&+&g^2[(G^+_{0}\Sigma)^{}_{n\alpha} (G^-_{0}\Sigma)^{}_{m\beta}] [(G^+_{0}\Sigma)^{}_{\alpha i} (G^-_{0}\Sigma)^{}_{\beta k}]\\
\label{eq:Wick3}
&+&g^2[(G^+_{0}\Sigma)^{}_{n\alpha} (G^-_{0}\Sigma)^{}_{\beta k}][(G^+_{0}\Sigma)^{}_{\alpha i}(G^-_{0}\Sigma)^{}_{m\beta}]\\
\nn
& +&  {\rm higher \; orders}.
\end{eqnarray}
Line~(\ref{eq:Wick2})
\begin{figure}[t]
\includegraphics[width=5.0cm]{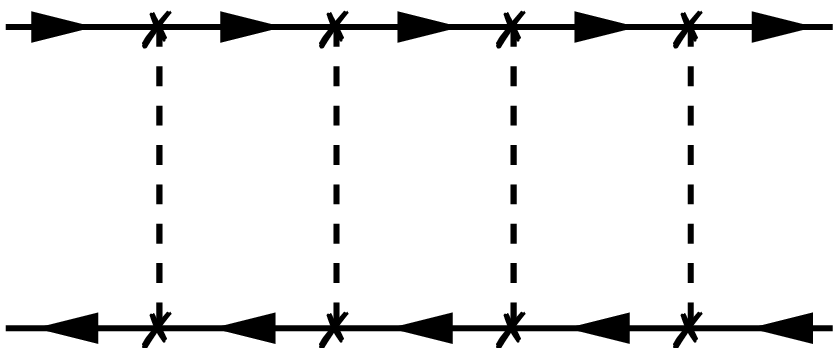}
\hspace{1cm}
\includegraphics[width=5.0cm]{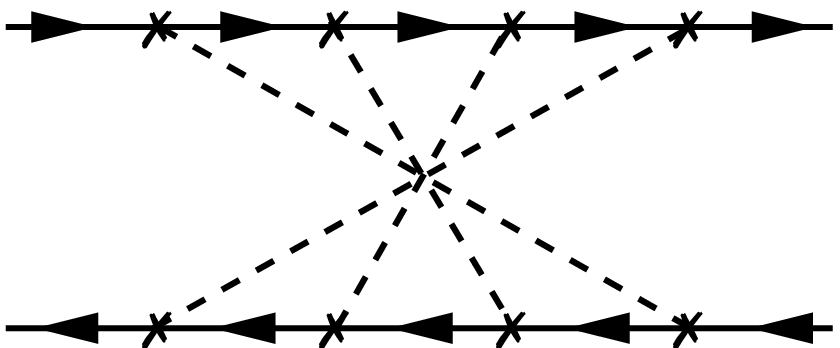}
\caption{A generic diagram from the ladder (left) and from the maximally crossed (right) class.}
\label{fig:Diagrams}
\end{figure}
represents a self--energy contribution. The total effect arising from summing all diagrams in this class can be accounted by the replacement of naked propagators by the renormalized ones, i.e.  $G^\pm_0\to\bar G^\pm$ via $\epsilon\to\epsilon+\eta$. The diagram class given by the complete geometric series 
\begin{eqnarray}
\nn
(1 - t)^{-1}_{rr^\prime|nm;ik} 
&=&
\delta^{}_{rr^\prime}\delta^{}_{ni}\delta^{}_{mk} +g\left[(G^+_{0}\Sigma)^{}_{ni}(G^-_{0}\Sigma)^{}_{mk} \right]^{}_{rr^\prime}\\ 
\nn
&+&
g^2\left[[(G^+_{0}\Sigma)^{}_{n\alpha} (G^-_{0}\Sigma)^{}_{m\beta}] [(G^+_{0}\Sigma)^{}_{\alpha i} (G^-_{0}\Sigma)^{}_{\beta k}]\right]^{}_{rr^\prime}+\cdots\\
\label{eq:ladders}
&=& \left[1 - g(\bar G^+\Sigma)(\bar G^-\Sigma) \right]^{-1}_{rr^\prime|nm;ik},
\end{eqnarray}
is called the ladder class. The corresponding diagrammatic representation is shown in Fig.~\ref{fig:Diagrams} on the left. For manipulations in   Line~(\ref{eq:Wick3}) we note that $(\bar G^-_{}\Sigma)^{}_{ab} = (\bar G^-_{}\Sigma)^T_{ba}$, where, the transposition operator $\rm T$ applies to all degrees of freedom, i.e. to the spatial ones as well. This class of diagrams is called the maximally crossed class and is shown in Fig.~\ref{fig:Diagrams} on the right. This series is incomplete with missing zero and first order terms. Introducing them we can perform the summation of this diagram class, obtaining
\begin{eqnarray}
\nn
(1-\tau)^{-1}_{rr^\prime|nk;im}
&=&\delta^{}_{rr^\prime}\delta^{}_{ni}\delta^{}_{km} + 
g\left[(\bar G^+_{}\Sigma)^{}_{ni}(\bar G^{-}\Sigma)^T_{km}
\right]^{}_{rr^\prime} \\
\nn
&+&g^2\left[[(\bar G^+_{}\Sigma)^{}_{n\alpha}(\bar G^{-}\Sigma)^T_{k\beta}]
[(\bar G^+_{}\Sigma)^{}_{\alpha i} (\bar G^{-}\Sigma)^T_{\beta m}] \right]^{}_{rr^\prime}+\cdots \\
&=& \left[1 - g(\bar G^+\Sigma)(\bar G^{-}\Sigma)^T\right]^{-1}_{rr^\prime|nk;im} .
\end{eqnarray}
Neglecting higher order diagrams we then obtain
\begin{eqnarray}
\nn
\langle G^+_{nj} G^-_{ml} \rangle  &\sim& 
\left(\left[1-g(\bar G^+\Sigma)(\bar G^-\Sigma)\right]^{-1}_{nm;ik} +
\left[1 - g(\bar G^+\Sigma)(\bar G^-\Sigma)^T\right]^{-1}_{nk;im} \right.\\
\label{eq:ATPGF}
&-&\left. \left[1 + g(\bar G^+\Sigma)(\bar G^{-}\Sigma)^T\right]^{}_{nk;im} 
\right)\bar G^+_{ij}\bar G^-_{kl},
\end{eqnarray}
where first term in the parenthesis denote the ladder channel matrix, second term the maximally crossed channel matrix and the remaining terms are introduced in order to avoid the double counting.  Addressing Eq.~(\ref{eq:ATPGF2}) it is clear, that none of the elements from the second line can develop a $\epsilon^{-2}$ singularity and thus disappear from the conductivity in the dc limit. Therefore we ignore them in further analysis.

\section{A pecularity of the reduced Kubo formula Eq.~(\ref{eq:WLKubo})} 
\label{app:WLKubo}

The reduced Kubo formula given in Eq.~(\ref{eq:WLKubo}) provides for a numer of wrong predictions for nodal electrons. For instance, it yields a finite dc conductivity for the clean system with a gap in the  spectrum. Indeed, at zero energy and in dc limit the Green's functions are simply 
\begin{equation}
\label{eq:MassivProp}
G^{}_{\pm}=[-i\nabla\cdot\sigma + m\sigma^{}_3]^{-1} 
\end{equation}
and hence Eq.~(\ref{eq:WLKubo}) yields 
\begin{equation}
\label{eq:weird}
 \sigma^{}_{\mu\mu} = \frac{e^2}{\hbar}{\rm Tr}\int\frac{d^2q}{(2\pi)^2}~\frac{[q\cdot\sigma+m\sigma^{}_3]\sigma^{}_\mu[q\cdot\sigma+m\sigma^{}_3]\sigma^{}_\mu}{[q^2+m^2]^2} \sim \frac{e^2}{2\pi\hbar}.
\end{equation}
This strange result is solely because of the neglected contributions of the type $G^{}_\pm\sigma_\mu G^{}_\pm\sigma^{}_\mu$, which annihilate it in the full Kubo formula: 
\begin{eqnarray}
\label{eq:FullKubo}
\sigma^{}_{\mu\mu} &\sim& {\rm Tr}\int\frac{d^2q}{(2\pi)^2} ~\left[G^{}_+(q) - G^{}_-(q)\right]v^{}_\mu\left[G^{}_-(q) - G^{}_+(q)\right]v^{}_\mu,
\end{eqnarray}
if Eq.~(\ref{eq:MassivProp}) is used. The equivalence of Eq.~(\ref{eq:FullKubo}) and Eq.~(\ref{eq:KuboGen2}) for translationally invariant systems is easily shown using the matrix identities
$\displaystyle G^{}_\pm(q) - G^{}_ \mp(q) \sim 2i\epsilon G^{}_\pm(q) G^{}_\mp(q),$ the definition of the velocity operator $v \sim \nabla^{}_{q}H$, and an integration by parts. Therefore while studying transport in nodal systems, the full Kubo formula has to be used.



\end{document}